\documentclass[aps,prd,showpacs,nofootinbib,preprint,tightenlines]{revtex4}
\def\be{\nopagebreak[3]\begin{equation}}
\def\ee{\end{equation}}
\def\ba{\nopagebreak[3]\begin{eqnarray}}
\def\ea{\end{eqnarray}}

\def\a{\alpha}

\def\l{\langle}
\def\r{\rangle}

\newcommand{\teta}{\rlap{\lower2ex\hbox{$\,\tilde{}$}}\eta{}}

\newcommand{\y}{\hat{y}}              % the quantum auxillary field y

\newcommand{\py}{{\hat{\pi}^{(y)}}}     % its conjugate momentum
\newcommand{\pyRI}{\hat{\pi}^{(y)R,I}}

\newcommand{\pf}{\hat{\pi}}           % its conjugate momentum
\newcommand{\fluc}[1]{(\Delta #1)^2}      % square of fluctuations
 
\newcommand{\cre}{\hat{a}^{\dagger}}    % creation

\newcommand{\ann}{\hat{a}}            % and annihilation operators
\begin{document}
\preprint{\vbox{\baselineskip=12pt \rightline{ICN-UNAM-05/01}
\rightline{gr-qc/0512013} }}
\title{Perspectives  on  Quantum Gravity Phenomenology}

\author{Daniel Sudarsky}
\email{sudarsky@nucleares.unam.mx} \affiliation{Instituto de
Ciencias Nucleares,
Universidad Nacional Aut\'onoma de M\'exico,\\
A. Postal 70-543, M\'exico D.F. 04510, M\'exico}

%\maketitle
\begin{abstract}
The idea that quantum gravity manifestations would be associated
with a violation of Lorentz invariance is very strongly bounded
and faces serious theoretical challenges. Other related ideas seem to be drowning in interpretational  quagmires.
 This leads us to
consider  alternative lines of thought for such phenomenological
search. We discuss the underlying viewpoints and briefly mention
their possible connections with other current theoretical ideas.

\end{abstract}

\pacs{04.60.-m, 04.60.Pp, 04.80.-y 11.30.Cp.}
 \maketitle

\section{Introduction}
The  search for a reconciliation of the view of space-time  as contemplated 
within the context of general relativity,
  and the principles of Quantum Theory has for most of its history been besieged by the  seemingly inescapable conclusion that 
no information  about the subject could, in practice,  be expected to emerge from the empirical realm. Nevertheless  the last few 
years  we have witnessed  a flare of interest on precisely this possibility. The change in outlook is due to the realization that 
in some simple scenarios,  some hypothetical manifestations of the effects presumably associated with Quantum Gravity  
(Q.G.) could become observable.  In those schemes  the Q.G. effects would be associated with a distortion of the microscopic symmetry
structure of space-time. These schemes can be divided into two subsets, in the first one assumes that the Lorentz symmetry
 is in fact broken and that Q.G. endows space-time with a preferential reference frame, a kind of  resurrected  ``Ether", while 
in a second class one introduces a modified Lorentz  and /or Poincar\'e structure without  invoking a preferential rest frame. Unfortunately these 
schemes suffer from some serious problems, which could in principle return us to the starting place with its bleak  outlook
 on possible phenomenological guidance in the  quest for a Quantum theory of Gravitation. 
  However,  now that the ``taboo" 
 about Q.G. Phenomenology has been broken, it seems appropriate to explore the notion in a larger context. In 
this spirit it  should be mentioned that the study of some possibilities that involve manifestations of Q.G. in 
the extended Poincar\'e algebra in conjunction with the Heisenberg  algebra  are already the object of intensive research
 \cite{Mendez}\cite{Chrysomalis}. In this paper we 
explore some options along two different lines of thought; the first one, motivated to  some degree by the ideas that where put forward 
in the context of the schemes that considered modification of the fundamental symmetries of space-time mentioned above,
  an that will be referred to as the ``space-time micro-structure 
 signature of Q.G." and the second one inspired by the ideas of  R. Penrose regarding the changes in standard Quantum Theory
 presumably tied with Gravitation,  and  an application to cosmology which we claim is already evidencing the need
 for some New Physics.

 This article is organized as follows: In section 2 we discuss some of the bounds that have been placed in  the models for  
Lorentz invariance violation associated with a preferential frame (Lorentz Invariance Violation  or L.I.V. in short)  
followed by
  what we regard as a devastating argument against this possibility.  In section 3 we briefly  discuss a series  problematic aspects of
 the ideas that have been  considered in the context of  modified fundamental  symmetry structures  
 of space-time without preferential frames. In section 4 we describe what seems to be the natural  descendent  of the schemes of L.I.V., 
and in section 5 we give short overview  of  R. Penorse's proposals for  a  Quantum Gravity induced ``collapse of the wave
  function" and follow it up with a recent  analysis indicating that something of that sort is needed if one wants to justify
 the arguments (and their predictive  success) leading from inflation to the birth of the cosmic structures.

\section{ Re-birth and Re-death of  Ether }

One immediate result that emerges once one starts thinking about putting together General  Relativity,  and Quantum
 mechanics 
is that there is a natural   scale with units of length that presumably signals the onset of the new physics: The Planck 
length $ l_{Pl}$.  The existence of such a fundamental length scale, when taken together with the well known special relativistic
  contraction  of lengths, has motivated  some researchers to consider the possibility that at the fundamental level the space-time 
structure, which in the quantum context is most naturally thought to be granular, determines by itself a local preferential frame, 
where the granularity has indeed the isotropic  scale $ l_{Pl}$ (in accordance with special relativity,  in other frames the scale 
 would be direction dependent).   Such situation, so the argument goes, would become manifest, most conspicuously, trough a modification
 in the dispersion relations for free particles, changing them into  something like \cite{Eather}:
\be
 E^2=( \vec P)^2 + m^2  + \xi E^3/M_{Pl}.
\label{DispRel}
\ee
 In  such  expression  the preferential  frame appears  indirectly: the equation being clearly non Lorentz Invariant, would be
 valid in, at most, one reference frame: the preferential rest frame.
 If we denote by $W^{\mu}$ the four velocity of the preferential  frame such expression can be written  in a Lorentz covariant  language:
\be
 P^{\mu} P_{\mu} = m^2  + \xi (W^{\mu} P_{\mu} )^3/M_{Pl}.
\label{CoDispRel}
\ee 
  As we will see, the existence of this four vector $W^{\mu}$, quite often hidden from the discussions, will have dramatic 
consequences. In fact any modification of the dispersion relations is conceivable only
 in  association with assumption of  the existence of new fundamental objects such as $W^\mu$.

 It is worthwhile   mentioning that, indeed, in the two most popular approaches to Quantum Gravity;  String Theory 
(see \cite{Strings}) and Loop Quantum Gravity (see \cite{Loops}), it has been argued that there is room for precisely the types of effects 
just discussed.

Most of the work along these lines have centered in direct searches for evidence of such modifications in the dispersion 
 relations, in  photons and electrons and  other elementary particles and interesting bounds have been obtained in the  
corresponding parameters  $\xi$ (which  in principle are taken as different for the various types of particles). For a comprehensive
   discussion of these results see the  review  article \cite{Mattingly}. 
 The central conclusion is that the bounds, extracted directly from these analysis, on
 the value of the parameters $\xi$, -- which were 
in principle expected to be of order one -- are in the range $10^{-4}- 10^{-9}$ for 
the most common particle species: photons, electrons, neutrons and protons  (i.e. quarks).  

The point we will focus on, and which represents the most devastating  argument against this approach can be traced to 
the following observation:  In the preceding discussion the underlying point of view has been that only very high energy particles
 are
 useful probes of these ideas, as seems to be indicated by the large suppression of order $E/M_{Pl}$ of the last term in 
equations \ref{DispRel}\&\ref{CoDispRel}
 above, relative to the dominant terms.  However  we must recall that  Quantum Field Theory teaches us that particles of all 
energies 
contribute to the virtual process that underscore all real  particle processes, and thus all processes  (even low energy ones) 
are influenced
 by the high energy modifications. In fact we should keep in mind that the dispersion  relations  correspond in fact to the location of
 the poles 
in propagators of the quantum field corresponding to the particle type in question\footnote{A note for the young reader: One should
 keep in mind that in working to combine special relativity and quantum mechanics one is taken quite generically into the realm of 
quantum field theory, and that particles cease to be
 fundamental entities and  are viewed instead as certain type of exited states of the quantum fields.}.
 These issues where first considered in \cite{Mayers},
 where the 
modifications where assumed to be associated with dimension 5 operators, naturally suppressed by the Planck mass scale.  
In that work the authors found that generically quantum loops lead to unsuppressed  
Lorentz Violating 
corrections in the propagators, which would imply  violations of Lorentz invariance of such magnitude as to be in blunt
 contradiction with observations. The dangerous expressions  originated from  quadratic diverging integrals
 -- and to a lesser extent also 
the linearly diverging ones -- that  would appear in connection with the otherwise  $M_{Pl}$ suppressed  effective Lagrangian
 operators  
that loops expansion generate.
An important detail
 that serves to illustrate a rather general point is the following : The  L.I.V. operators in the original Lagrangian,  all 
contained the triple product 
 $W^{\mu} W^{\nu} W^{\rho}$.  The dangerous integrals are of the form 
$\int (k_{\mu}k_{\nu} /k^2)d^4k \approx \eta_{\mu \nu}$. 
 Within the
 effective field theory approach the dangerous integrals would need to be considered as having a cut-off at a the large mass 
 scale  $\Lambda$ 
and thus would be proportional to $\Lambda^2$. 
 When taken together within the structure of the original L.I.V. operators  one  would end 
up with  an effective term of order $ \Lambda ^2 / M_{Pl}$, and structure product 
 $ W^{\mu} W^{\nu} W^{\rho}\eta_{\mu \nu} =- W^{\rho}$,
 which would thus lead clearly to a situation with an exceedingly large  L.I.V.. The  authors  of  this work then attempt to evade the
 disastrous conclusions, by
 considering an ``add hoc" proposal to get rid of 
these dangerous terms: to replace in the  structure  of the dimension 5 operators every occurrence of
the triple product of $ W $'s by the tensor $C^{\mu\nu \rho} =W^{\mu} W^{\nu} W^{\rho} -
(1/6)[\eta^{\mu \nu}W^{\rho} +  \eta^{\mu \rho}W^{\nu} + \eta^{ \nu \rho}W^{\mu}] $ which has the property that 
it vanishes upon contraction  of any 
two induces with the metric tensor  $\eta $. 
The odd thing about this is the following: The dangerous integrals are normally argued to give a result proportional
 to $\eta$ relying on Lorentz
 Symmetry arguments. Thus one's position has effectively become the reliance  on a symmetry  -- that is assumed to be broken -- to
 ensure 
that the loop corrections do not generate the operators that would break it too badly.  This sounds very dangerous.  
Indeed  it has
 been shown in \cite{AlexI} that, upon the consideration of higher order diagrams, the scheme proposed in \cite{Mayers} 
fails, and one ends up with the 
large L.I.V. one was trying 
to avoid.  

In fact such  problems can be seen to be rather generic, in the following sense: Let us take seriously 
 the motivational  arguments
 mentioned above, which  lead to the  suspicion  that Quantum Gravity might be associated with a breakdown of  Lorentz
 invariance,  
 and let us consider that at the fundamental scale space-time has a discrete structure  characterized by the Planck length, and
 take
 this to indicate an underlying granular structure  in space-time, with such characteristic  length scale,  as seen from its proper  reference 
frame\footnote{ Such proper
 reference frame could be thought of as that in which the granular structure is maximally isotropic.} which we will call the 
{\it fundamental frame}. In that case,  the consistent  treatment  of  the theory should include a provision indicating that 
there is a bound, with a fixed and  specific value, on the physical wavelength of excitations   as seen in the fundamental frame.
 Therefore, the quantum theory should not contain 
the corresponding excitations, either as real or as virtual particles. 
Similar considerations have in fact been made in proposing a saturation of the
de-Broglie  wave  length at the Planck length as the momentum goes to infinity \cite{saturation}.
 Thus  {\it every theory} which we now consider as candidate 
to  be a fundamental theory, should be regarded instead, as merely an effective theory and it should include a momentum cutoff, eliminating 
 the unphysical excitations.  That is,  we must for instance take the standard model of particle physics and
impose on it a cut-off on the particle's  3-momentum as seen in the 
fundamental frame\footnote{ One  could of course argue that the standard model is in reality also an effective theory,
and that for some  unspecified reasons
 such arguments should not
 apply to it directly but to some more fundamental and  yet unknown theory, but then, one would have to give up 
the argument that motivates 
the search for these quantum gravity effects using probes and interactions that are described in terms of this theory.}.
This feature would in principle have to be combined with other features of the effective theory, such as the change
 in the propagators of 
particles which  would correspond to the modified dispersion relations.  However  we will see that,  as shown first in \cite{Collins},
 the effect of the frame
 dependent cut-off by itself is disastrous. In order to do this we  consider the full propagator of a scalar particle in
 Yukawa  theory.  We 
 focus on this theory,  in order to illustrate the  main point,   because of its simplicity and because of 
the fact that is part of the standard model of particle physics,
 and thus we can rely
 on the wealth of knowledge about its phenomenology to confront with  the consequences of the aforementioned ideas.

The theory is  defined by the Lagrangian density:
\begin{eqnarray}
\label{eq:L.Yukawa}
  {\cal L} &=& {1 \over 2} (\partial\phi)^2 - \frac{m_0^2}{2} \phi^2
             + \bar\psi (i\gamma^\mu\partial_\mu - M_0) \psi
             + g_0 \phi\bar\psi\psi.
\end{eqnarray}
We  introduce next  the  cutoff on spatial
momenta in the fundamental frame.  Of course this is not a realistic
model of Planck-scale granularity.  However it does represent a field
theory in which the basic Lagrangian gives Lorentz-invariant
dispersion relations for low-energy classical modes, and in which
there is a Planck-scale cutoff that is bound to a particular frame.

Therefore fermion bare  propagators are  modified according to
\be
\frac{i}{\gamma^\mu p_\mu -m_0 +i\epsilon} \rightarrow 
\frac{i f(|\vec p|/\Lambda)}{\gamma^\mu p_\mu -m_0 +\Delta(|\vec p|/\lambda) +i\epsilon}, 
\label{propFermions}
\ee
 and similarly the scalar bare propagators are modified according to
\be
\frac{i}{p^2 -M_0^2 +i\epsilon} \rightarrow \frac{i \tilde f(|\vec p|/\Lambda)}{p^2 -M_0^2 +\tilde\Delta(|\vec p|/\lambda) +i\epsilon}.
\label{propScalar}
\ee
The requirement on the  functions  $f(x)$ and $\tilde f(x)$ which specify the cut-off is that they go to $1$ as $x\to 0$
 to reproduce  Lorentz Invariant low energy behavior and that they go
 to zero as $ x\to\infty$. The functions $\Delta $ and $\tilde\Delta$  would be specified by concrete proposals for the modified dispersion relations.
  We concentrate in  examining  the effect of the cutoff by itself, and thus the changes in 
the bare dispersion relations will be ignored.
Let us  consider the full
propagator of the scalar field, and more specifically
 $\Pi(p)$  its self-energy\footnote{In perturbation theory,
  $\Pi(p)$ is the sum over one-particle-irreducible two point graphs
  for the scalar field. }.  The
parameter $\Lambda$ is of order the Planck scale.  Our choice that the
cutoff function depends only on the size of the 3-momentum is
for simplicity of calculation, and in fact simple changes in this choice do not change the main result \cite{Alexis}.

The one-loop approximation to the self-energy $\Pi(p)$ is given by the
 a standard  ``sausage" Feynman diagram.  
We wish to investigate its properties when the momentum $p^\mu$ and the
mass $m$ are much less than the cutoff $\Lambda$.  Thus we make the  customary  Taylor expansion
of $\Pi$ about $p=0$ and  obtain
\be
   \Pi(p) =A + p^2 B +
             p^\mu p^\nu W_\mu W_\nu \tilde\xi
     + \Pi^{\rm (LI)}(p^2) + {\cal O}{(p^4/\Lambda^2)} .
\ee
Here $W_{\mu}$ is the 4-velocity of the preferential frame, which appearance can be traced to 
eqs. \ref{propFermions} and \ref{propScalar}  where $|\vec p| =\sqrt{(\eta_{\mu\nu} +W_\mu W_\nu)p^\mu p^\nu}$. 
$p^2=p^\mu p^\nu\eta_{\mu\nu}$, with $\eta_{\mu\nu}$ being the space-time
metric. The coefficients $A$ and $B$ correspond to the usual
Lorentz-invariant mass and wave function renormalization. The fourth term $\Pi^{\rm (LI)}(p^2)$ is Lorentz-invariant. 
The third term however is clearly Lorentz violating. The
coefficient $ \tilde \xi$ is independent of $\Lambda$, and in fact
   explicit calculations give:
\begin{equation}
\label{vani}
 \tilde  \xi = \frac{g^2} {6\pi^2}
       \left[ 1 + 2 \int \limits_0^{\infty}{dx} x f'(x)^2 \right] .
\end{equation}
Although this term depends on the details of the function $f$ which
models the microscopic quantum gravity effects, it is positive
definite.  Quantitatively the corresponding Lorentz violation is of
order the square of the coupling, rather than being power-suppressed.
One might want to treat this term as a renormalization of the
space-time metric tensor however, there are many fields in
the standard model that differ by the sizes of their couplings.  Hence one way to describe the effect 
would be to say that each of these fields
 sees a different metric tensor and thus has a different limiting velocity.  On the other hand, the limits on  the differences in limiting
 velocities for different particle species are -- even using only analysis that predate the latest round of studies -- quite stringent \cite{Weinberg} at the 
level $10^{-20}$
while  the expected value of such differences  is about $10^{-2}$ at the least given the values of the standard model 
coupling constants.
  Thus in the absence of a  mechanism that would prevent this large Lorentz violations,  while preserving some small ones, 
the ideas underlying the  proposals 
  discussed at the start of this section, would  seem to be rather untenable.  Recently Pospelov \cite{Pospelov} has argued
 that supersymmetry might provide such mechanism. However he notes that the supersymmetry algebra contains  
the Lorentz Algebra, and thus it would seem problematic to argue that the latter is broken but the former is at work.
  He notes nevertheless
that the protection from large Lorentz violations would  work even if only the translation  subgroup was unbroken.
 Here, we point out that 
Space-time granularity would break precisely such subgroup. 

We should emphasize that while the idea that the Planck length
might  be some sort of  ``minimum measurable length", seem to put into question
 the range of validity of  Lorentz invariance, 
 as shown in \cite{carlo}, the existence
of a minimum measurable length does not of itself imply that local
Lorentz invariance is violated any more than the discreteness of
the \emph{eigenvalues} of the angular momentum operators implies
violation of rotational invariance in ordinary quantum mechanics. 
On a similar note,
 the work in \cite{Rafael}
illustrates the point that a discrete structure
 of space-time does not by itself imply the existence of a preferential
 rest frame or the violation of Lorentz invariance.

We conclude that at present the theoretical ideas that pretend to connect a space-time granularity of quantum gravitational
 origin, 
with a violation of Lorentz invariance  seem to be in serious trouble, to say the least.

\section{ If it ainÕt broken why not try  bending it?.}

This title has perhaps an  exceedingly  negative  connotation, 
 and fairness requires that the reader be warned  that  at this point, it can not be argued
 that these approaches are unviable.  However,  given the tremendous difficulties that such schemes seem
 to face, in particular as regards to the physical interpretation  that one is to give to the mathematical  
structures,  and which will be briefly discussed below, I can only give my own  personal pessimistic 
 outlook for this line of 
thought.  Nevertheless, I shall point to a particular  deviation  that seems to me more promising not only  because it is simpler,
 but rather because, not only is it grounded in a  rigorous mathematical foundation, but is based on a method that 
is successful in what can be considered as similar instances.

This second point of view towards Quantum Gravity Phenomenology is based on the idea that Lorentz 
 Invariance might not be broken and that there would be therefore no  preferential frame at all, but that instead, the local
 geometry of space-time would exhibit departures  from that described by special relativity.  The options that have been
 considered
 can be classified as tied to the notions: 1) that the Lorentz algebra might be replaced by  some sort of nonlinear mathematical structure,
  2) that  the Lorentz algebra, in fact the full  Poincar\'e algebra
might be unified   with the Heisenberg algebra  (i.e. including  space-time  coordinates) 
and  then  modified, 
and 3) that the space-time structure  {\it  in itself }might become 
non-commutative . 

To go into the detailed way in which  problems appear in each one of these alternatives would  be far 
beyond the scope of this paper. However  the main problems will  be mentioned for the benefit of the reader.

Consider first the scenarios where the Lorentz  algebra is supposed to appear in  a nonlinear form, i.e. where the commutators are not longer
 linear functions of the generators.  
It has been  shown that in the cases that have been studied \cite{redefinitions} one can perform a nonlinear redefinition of the generators 
in which the algebra  takes  again  a linear  Lie algebra form
 (if one adds the central generator and then  proceeds as in \cite{NonLinear-linear}). 
 One then  takes the view that those
 are not the variables that one measures. 
Then the issue becomes
 what are the variables that we measure?, and this takes us to the assumptions underlying the way we built our detectors and other
 devices that measure energy momentum etc.. Here we note that the conservation of  energy-momentum is one of the basic principles
  in which we  base our measurements of these quantities, our design of the apparatuses to measure them, and that  this fact  is a
 particularly important aspect  of such measurements for high energy particles.
  Thus, the issue  becomes what  are
 the quantities that are conserved?, and this takes us to the issue of how one obtains  total energy and total momentum for
 composite 
objects, which  is essentially linked to the selection of the co-product for the  algebra  generators.

If one wants something different from the standard situation one would need  a 
nontrivial  (non-primitive) co-product \cite{coproduct}.
 The cases that have been analyzed are based on the selection of asymmetric (or non-commutative)  co-products, where,  say, the total energy of 
a pair of particles depends on  
the way we order them calling them first or second. Here one faces a very serious problem because there 
does not seem to  exist a canonical
 recipe for deciding in each specific situation,  which order to take, i.e. given two particles in a collider,  which one is  
called the first
 and which is second affects their total energy and momentum, a clearly disastrous situation, as one would not know how 
to proceed.
Thus one would lack an interpretation  
scheme for the formalism, 
a fact that makes it impossible to use, at least, for phenomenology.

In fact even if one chooses  a symmetric  but nontrivial   co-product one faces the so called ``spectator problem" where the system of two
 particles would transform 
differently if 
considered as a subsystem of, say a three particle system, that when considered by itself. In that case one would not know in 
principle how to proceed, 
as even a particle in a remote region of the universe, which happened to be in the regime where the non-linearity becomes important,
 would affect the physics 
of, say, a 
scattering process at  a Fermilab.
It thus seems that any recourse to a nontrivial co-product puts us in an essentially untenable situation. Let us note that some
 of the problems that these proposals
 face have  been pointed out before \cite{problems}.

The second option  would start from the requirement that one maintains a Lie Algebra structure and a trivial co-product, and
 modify  only the Lie Algebra  structure constants.
Within this set of ideas, the most promising  ones seem to be those that arise  from considerations of  algebraic 
 stability  applied to the Poincar\'e-Heisenberg algebra, \cite{Mendez}, \cite{Chrysomalis}. It is noteworthy that such considerations would
 take one directly from the  Galilean algebra, to the Lorentz algebra,  and from the commutative  algebra of functions over phase space,
 to the Heisenberg algebra \cite{Flato}.  
Unfortunately, and as clearly noted by the authors of \cite{Chrysomalis}
 this approach also suffers from interpretational difficulties, connected to the fact that  for
 composite systems the position operators 
can not be reasonably expected to be additive. 
 In other words, in such schemes one is asked to consider nonstandard 
commutators involving the 4-position and 4-momentum 
operators  and a new central operator,  while the Lorentz sector 
remains untouched.  In devising an experiment we need to have an 
unambiguous identification of these objects with the quantities one measures.  
It seems rather clear that the objects we would call the position operators in these schemes
can not be identified with the actual position of objects that are obtained during 
measurements\footnote{The issue is the following:  The position operator is not expected to de additive
 (to find the position of an hydrogen atom one does not add the position of the proton wit  that of the electron), 
while the symmetry of the construction would require similar coproducts for position and momentum operators (the momentum of composite
 systems as usual being additive). 
On the other hand if the coproduct is nontrivial, the issue would be how to deal with the position operators
 for composite objects.}, and then  one is at a loss as to what  can be an actual test of the scheme.  
It is  a fact that the momentum operators do not seem, at first sight, to suffer from the
same problems that afflict the position operators but the issue of 
what  objects  can they in fact be associated with,  seems a bit confusing as the momentum operators are intimately connected
 to the position operators as they both trace their origin to  conjugated  pair of variables.  Thus it seems that one 
could not associate such momentum operators to objects to which one can not associate  the corresponding position operators.  
These points are not made to  suggest that 
the scheme is unviable,  as I do not think it is, but  rather to stress the fact that, the 
hope for its applicability  lies in a profound interpretational  analysis that would clarify the
 status of ``4- position 
observables" and connect them to the objects found in the real world.

The third set of ideas, the so called non-commutative geometry program starts usually with the postulate that the Minkoswki
 coordinates do not commute but the commutators are functions of these coordinates 
themselves and not of any other generators. Thus one assumes
that they  satisfy a fundamental commutation relation such as
\be
[ x^\mu, x^\nu ] = i \theta^{\mu \nu},
\ee
 where $\theta^{\mu \nu}$ is a fundamental  antisymmetric c-number tensor.  Here we note  that, if taken at face value 
and using it  without  further modifications\footnote{It is possible  of course,  to bring in a more elaborated structure to
 remove the problems of a naive interpretation.  In fact in the  more  methodical  formulations many other objects
 become non-commutative,  including for instance  entries of the matrices representing
the generators of the Lorentz 
transformations, resulting in new  notions of invariance and  new types of invariant tensors \cite{NonComm-LT} }, the problem is  
 that there is no  Lorentz 
invariant  antisymmetric tensor of rank 2. This point should not confused with  the covariance of  antisymmetric tensors such as the
electromagnetic   field strength $F^{\mu \nu}$.  The issue is of course, that once we write the specific numerical
 proposal for 
the matrix $\theta^{\mu \nu}$, such specific value can be associated at most with one specific  reference frame, and then 
the issue is: which one?
In other words, while in the case of  $F^{\mu \nu}$ its specific values in  a given frame  and a given physical situation are  
determined by the  field 
equations of motion and the boundary conditions (the latter of course have different specific values in the various frames), in the case
 of a fixed fundamental object associated with the space-time structure such as 
$ \theta^{\mu \nu}$ those  elements are not available.  Therefore any specific recipe for  $ \theta^{\mu \nu}$ could only be
done in association with one specific  frame, and furthermore it would
imply the existence of a preferential
frame where the corresponding matrix with specific values  has a particularly simple form. For instance if at all possible it will
be only  
only in one particular  frame that we could say that the $\theta^{0 i}$ components are all zero.
  Thus this scheme incorporates the selection of preferential frames, and is thus, as pointed out in \cite{Collins},
 susceptible to the same problems we encountered  in
section II.  Of course one should stress that it is conceivable that a scheme can be constructed that would overcome these difficulties 
and in this regard its worthwhile 
to mention the proposals in \cite{Paolo}.  

Other proposals one finds in the literature,  start by writing:
\be
[x^0, x^i] = i\lambda x^i,\qquad or,  \qquad [x^\mu, x^\nu] =f^{\mu \nu}_\rho x^\rho,  
%or \qquad, x^\mu x^\nu=R^{\mu nu}_{\rho\sigma}x^\rho x^\sigma .
\ee
 These schemes in my view  suffer from another
 serious problem: LetÕs  focus on the first proposal:  If we pretend to view this within the interpretational framework of quantum mechanics,
the corresponding uncertainty relations indicate that one could not find a simultaneous eigenvector  of
 $x^0$ and $ x^i$, except when the corresponding eigenvalue  of $ x^i$ happens to be zero. Thus one can not localize an
 event both in space and in time (taking as we said the interpretational framework directly from QM)
 unless that event is at the origin of the coordinates.  The issue is then:  Where is the origin of coordinates?
 It would be clearly unphysical to state that certain points in space become physically differentiated just because we chose 
them as the origin of coordinates. 
  It is clear that the second scheme 
% and third schemes
 suffers from similar problems:  If the objects $x^\mu$  have any relation whatsoever with the 
coordinates we measure  it is clear that the non-commutativity becomes larger for larger values of the coordinates 
 ( the same can be said  for eigenvalues, expectation values, for the corresponding operators, and in fact for any adjudication of some
  real values to these objects).
 In other words the effect decreases as we approach the origin of the coordinates.  But where on the universe is this point or region?
The only option seem to be to change the interpretational scheme, but then we find ourselves
 again in a similar conundrum as that of  the first direction we explored in this section.
  
It is of course possible that these problems might be overcome but at this point it is fair to say that the situation if far from clear.
 For further reading on Quantum Field theory in
 non-commutative  space-times see \cite{QFTNCST}.

\section{ What might be There?}
\noindent 
In view of the hardships one encounters in trying to reconcile the  naive idea of a
 granular structure of space-time -- which would naturally 
be associated with a preferential reference frame where the granularity takes
 say the most symmetric form -- with the tight phenomenological bounds
that have been obtained  and with the clear expectations from Quantum Field Theory, that 
the effects of such granular structure would 
be  only lightly suppressed, one is lead  to consider   more subtle possibilities. 
 In this section we will discuss an alternative way in which a granular structure of space-time might appear, and 
which would be immune
 from the previous considerations while still, in principle, susceptible to a phenomenological study. 
The idea will be discussed in 
 a rather heuristic 
way  and it is fair to say that there is at this point no concrete realization of the proposal. However one can as usual employ
 the symmetry principles  to restrict the possible phenomenological manifestations. These ideas have been studied 
first in \cite{NewQGP} 

 We have at this point no real  good  geometrical picture of how a granularity might be 
associated to space-time while strictly preserving the Lorentz, and Poincar\'e symmetries. 
One can point however to the Poset program \cite{Rafael} as one that seems to embody such scheme,
 however we will not  at this point  commit to any such specific proposal.  Instead we seek  guidance in analogies with 
some simple ideas  from solid state physics.
Thus, we consider the case of a crystal, and note  that when a large
crystal has the {\it same} symmetry (say cubic)  of the fundamental
crystal, one could expect no deviations from fully cubic symmetry,
as a result of the discrete nature of the fundamental building
blocks.  In fact one would not expect in such situation that the discrete structure 
of the crystal could be revealed at the macroscopic level by any deviation from precise cubic symmetry. 
 The discrete structure might be studied, of course, but NOT by looking at deviations from such symmetry.
However if one considers  a macroscopic crystal whose
global form is not compatible with the structure of the
fundamental crystals, say hexagonal,  the surface will necessarily
include some roughness, and thus a manifestation of the granular
structure, would occur through a breakdown of the  exact hexagonal
symmetry.

 Our ideas will be guided by the simple picture above, which will be transported {\it mutatis mutandis}
from the crystal and the cubic symmetry to the space-time and the Lorentz symmetry. 
Thus, we will start by assuming
that the underlying
symmetry of the fundamental structure  of space-time is itself the Lorentz
Symmetry,  which would naturally leads us to expect no  violation of the symmetry at the macroscopic level
when the space-time is macroscopically Lorentz invariant.
Thus,  the large scale Lorentz
Symmetry is protected by the symmetry of the fundamental granular
structure.

 Thus in  a region  of space-time normally
considered as well approximated by Minkowski metric, the 
granular structure of the quantum space-time would not become manifest through
the breakdown of its symmetry. 
 However, and following with our solid state analogy, we are lead to consider the  situation 
 in which the macroscopic space-time  is
not fully compatible with the symmetry of its basic constituents.

The main point is then, that in the
event of a  failure of the  space-time to be exactly Minkowski in
an open domain,  the underlying granular structure of quantum
gravity origin, could become manifest, affecting the propagation
of  the various matter fields. Such 
situation should thus involve the Riemann tensor, which is known
to precisely describe the failure of a space-time to be Minkowski
over an open region. Thus the
non-vanishing of Riemann would correspond to the macroscopic
description of the situation where the microscopic structure of
space-time might become manifest. Moreover, we can expect, due to
the implicit correspondence of the macroscopic description with
the more fundamental one, that the Riemann tensor  would also
indicate the space-time directions with which  the sought effects
would be associated.

This  selection of special space-time directions, embodies a certain analogy within the current approach, to
the global selection of a preferential reference frame that was
implicit in the  schemes 
 towards Quantum Gravity Phenomenology  described in section  II.
%\ref{}.

 With this ideas in mind we turn now to proposing the corresponding phenomenology. 
That would imply the consideration of  an effective  description in the way that the Riemannian curvature
could affect, in a nontrivial manner, the propagation of matter fields. Thus we need to consider the Lagrangian terms representing such couplings.
 
Before we do so, we  
recall  that the Ricci tensor represents that part of the Riemann
tensor which, at least on shell, is locally determined by the
energy momentum of matter at the events of interest. Thus the
coupling of matter to the Ricci tensor part of the Riemann tensor
would, at the phenomenological level, reflect a sort of pointwise
self interaction of matter that would amount to a locally defined
renormalization of the usual phenomenological terms such as a the mass
or the  kinetic terms in the Lagrangian.

 However we are interested in
the underlying structure of space-time rather that the self
interaction of matter. Thus we would need to ignore the aspects
that   encode the latter, which in our case would
corresponds  to all Lagrangian terms containing   the Ricci tensor,  coupled to
matter fields. The remainder of the Riemann tensor, i.e. the Weyl tensor, can thus be thought,
to reflect the aspects of  the local structure of space-time associated
solely with the gravitational degrees of freedom.

Therefore we are lead to consider
the coupling of the Weyl tensor with the matter fields. We note that 
 in the absence of gravitational waves,
the Weyl tensor is also connected with the nearby ``matter
sources" but such connection involves the propagation of their
influence through the space-time and thus the structure of the
latter would be playing a central role in the way the influences
become manifest. In this sense the Weyl tensor reflects the
``non-local effects" of the matter in contrast with the Ricci
tensor or curvature scalar that are determinable from the latter
in a completely local way.

We  further assume observer
covariance and the absence of globally defined non-dynamical
tensor fields.
 We are interested in the minimally suppressed terms, those that are only suppressed by the first power of 
$ M_{Planck}$, which would naturally   correspond to the dimension
5 operators, while  considering the  coupling of the 
fundamental fields of the standard model, bosons
and fermions, to the Weyl tensor.  Using the fact that the Weyl tensor, as the Riemann tensor, has mass
dimension 2, while the fermions have mass dimension 3/2 and the bosons
have mass dimension 1, one can show that  there are no non-vanishing
 dimension five operators coupling the Weyl tensor to the fields of the standard model \cite{NewQGP}. 
 Thus one can either take this as an indication  that  the effects one is looking for are more strongly suppressed 
or search for somehow more indirect approaches.
 In \cite{NewQGP}  we take the latter approach and consider the following schema.

One considers the Weyl tensor viewed as a tensor
of type $(2,2)$ as a  mapping from the space of antisymmetric
tensors of type $ (0,2)$, $\cal S$  into itself. As is well known
the space-time metric  endows the six dimensional vector space
$\cal S$ with a pseudo-Riemannian metric of signature $(+++---)$.
%\cite{wald}. 
Then the Weyl tensor is a symmetric operator on this
space ${\cal S}$ , which can therefore be diagonalized, and thus
has a complete set of eigenvectors (which are however not
necessarily orthogonal). We will assume for simplicity, that all eigenvalues are
different, and consider only the eigenvectors
$\Xi^{(i)}$ corresponding to non-vanishing eigenvalues
$\lambda^{(i)}$, by fixing the normalization of these eigenvectors
to be $\pm 1$ (also drop the null eigenvectors). Next we use the antisymmetric tensors
$\Xi^{(i)}_{\mu \nu}$ and their associated eigenvalues
$\lambda^{(i)}$ to construct the types of Lagrangian terms we are
interested in. Finally we look for terms
linear in these objects, and recalling that the eigenvalues
$\lambda^{(i)} $ have the dimension of the Riemann tensor, we have
the least possible suppressions in each sector as follows:  In the
scalar  sector there is in fact no  candidate of dimension 5 or 6 for such term.
In the vector boson sector, taking into account the requirements of
gauge invariance,  we are lead to a dimension 6 term
\be {\cal L}_{\rm m}=\frac{\xi }{M_{\rm Pl}^2}\,   \sum_{i}\,
\lambda^{i}\, \Xi^{(i)}_{\mu \nu} \,{\rm Tr}\,( F^{\mu}_{\rho} F^
{\rho\nu} )\, .
 \ee
It is worthwhile pointing out that in a  purely $ U(1)$ sector one can write a dimension 4 term,
\be
 {\cal L}_{\rm m}= \xi  \sum_{i}\,
\lambda^{i}\, \Xi^{(i)}_{\mu \nu} \, F^
{\mu\nu} \, .
 \ee
 This is  an unsuppressed term, which  is rather surprising, however given that  such term can not be
 written in the case of non-abelian gauge fields, together with fact that 
in the standard model the $U(1)$ sector mixes with the $SU(2)$ sector suggest that such terms should be absent.
 We have no tighter argument 
regarding this possibility at this point, but we will not consider it any further  in view of the last observation.

 Finally, in the fermion  sector we have a term,
\be {\cal L}_{\psi}=\frac{\xi }{M_{\rm Pl}} \, \sum_{i}\,
\lambda^{i} \,\Xi^{(i)}_{\mu \nu} \,\bar\Psi
\gamma^{\mu}\gamma^{\nu}\,\Psi\, .\label{neta}
 \ee

Thus the fermions seem to provide the most promising probes, which seems a
fortunate situation, in this scheme.

  One could also consider
coupling
  directly a scalar made out of the standard model fields to an
  appropriate power of a
  scalar constructed out of the Weyl tensor such as
  $(W_{\mu\nu\rho\sigma}W^{\mu\nu\rho\sigma})^{1/2}$.  This
  proposal,  departs slightly from the  spirit of the suggestion the space-time structure would naturally
  and locally select preferential space-time  directions. The lack this feature would tend to make the
  effects, in principle 
  much harder to detect experimentally. On the other hand this
  line opens the way to consider,
   effects that would not be
  suppressed by  $M_{Planck}$ at all, such as
  \be
 {\cal L}_{\psi}=
(W_{\mu\nu\rho\sigma}W^{\mu\nu\rho\sigma})^{1/4}\bar\Psi\Psi.
    \label{neta2}
 \ee
 These type of terms  in which the space-time structure appears only as a scalar coupled 
to the matter fields would correspond
to a space-time dependence of mass or coupling constants,  controlled by local curvature. 
As we mentioned, the fact that they exhibit no 
particular signature, 
would tend to make  the related effects,  very difficult to probe.

Regarding phenomenology one should  thus, concentrate clearly in the fermion sector
as  leading to the most promisingly observable
effects.

Before continuing we write again the corresponding Lagrangian term, taking  now into
account, a possible flavor dependence, which could be thought to arise from the detailed way the 
 different fields interact with
the virtual excitations that intimately probe the underlying space-time  structure. Thus we consider:
\be {\cal L}^{(2)}_{f}=\sum_{a}\frac{\xi_a}{M_{\rm Pl}}\,
\sum_{i}\, \lambda^{i}
 \, \Xi^{(i)}_{\mu \nu}
 \, \bar\Psi_a  \gamma^{\mu}\gamma^{\nu}\Psi_a\, .\label{10}
\ee
where $a$ denotes flavor.  Next we note that we have in principle
the same types of effects that have  been considered in  the
Standard Model Extension (SME) \cite{SME} but only with terms of
the form $-1/2 H_{\mu \nu}\, \bar\Psi \sigma ^{\mu \nu} \Psi$.
Moreover, here  the tensor $H_{\mu \nu}$ must be identified with $
-\frac{2\xi }{M_{\rm Pl}} \sum_{i} \lambda^{i}\, \Xi^{(i)}_{\mu
\nu}$, and thus has a predetermined space-time dependence dictated by the
surrounding gravitational environment.
Therefore, special  care has to be taken when comparing  different experiments at different sites\footnote{This is reminiscent of the
 Situation  encountered with the studies of  the ``Fifth Force" proposals\cite{Ephraim}.},
by taking into account the differences in the surrounding
environment that leads to variable values of the relevant
curvature related tensors. 

Finally we briefly comment on the related phenomenology:  The relevant experiments must be  associated with both,
relative large gravitational tidal effects (indicating large
curvature)  in the local environment together with probes  involving polarized matter as the explicit
appearance of the Dirac matrix $[\gamma^\mu, \gamma^{\nu}]$
indicates.   Both conditions  seem from the onset difficult to achieve, and to control. Polarized matter 
is usually highly magnetic  and thus electromagnetic disturbance  would  need to be controlled to a very high degree as
they would tend to obscure any possible effects.  Gravitational field gradients are usually exceedingly  small  on Earth and even in
 the solar system.

 Thus, neutrinos crossing regions of large
curvature, seem like  very good candidates to be studied in this context. We note in particular
 that a term of the sort we are considering could lead
to neutrino oscillations even if they are massless, in close
analogy with the ideas exposed in \cite{neutrinos}.

Next we note that  the terms  in question do not violate CPT so
that that particular phenomenological avenue is closed. On the
other hand other discrete symmetries, particularly CP could,
depending on the environment and state of motion of the probes seem to be
open channels for investigation.

In this light it would be very interesting to  consider the Neutral Kaon system
where, as in the fifth force scenario, one would look for energy
dependence of the system's parameters \cite{MyThesis}. On the other hand, as we 
mentioned before one expects the
that the useful probes would involve polarized matter which would seem to rule out the usefulness of  Neutral Kaons. 
However one should consider other particles, such as neutrinos that  might combine, some sort of
 flavor oscillation with a nontrivial
 polarization structure. 

 All these ideas are of course in need of a much more detailed study. 

We end this section by pointing out a quite  different proposal
 regarding possible manifestations of Quantum Gravity:  the possibility that  a underlying discrete structure of time,
 would lead to a 
fundamental decoherence  in 
quantum mechanics considered  in \cite{GIDecoh}.  This idea is quite intriguing and in fact is closer in spirit to the ideas
and proposals  that we address in the next section.

\section{What seems to be There}

  This might sound like a strange title, as it indicates that there 
is  in fact some sort of evidence  for a manifestation of Quantum Gravity.  We will argue that indeed there is something  out there
that requires  new physics for its understanding.  It is of course not at all  clear that the problem we will discuss
 should be related to Quantum Gravity, but
since that is the only sphere of fundamental physics for which  we have so far failed to find 
 a satisfactory  conceptual understanding\footnote{There are of course many open issues in fundamental 
understanding of physics that are not in principle connected with the issue of quantum gravity, however  
it is only in this latter field that the problems seem to be connected with deep conceptual  issues and where 
 one can envision the possibility that their resolution might require a fundamental 
change of paradigm, as the  would be the case if we find  we  must modify the laws of quantum mechanics.} we find quite
 natural to associate the two. In fact the ideas of Penrose  regarding the fundamental
 changes,  that he argues\cite{Penrose}, are needed in  Quantum Mechanics and their connection to quantum Gravity, are a 
inspirational precedent for the analysis first reported in \cite{InflationUS}.

  There are for instance  lingering interpretational problems in quantum mechanics, in particular in connection with
the measurement problem. For instance, and  as it is often  emphasized by R. Penrose,
we have in the Copenhagen interpretation, and in fact in any practical application of the theory, two quite different evolution
 processes: the U process or unitary evolution applied when systems 
are not subjected to a measurement, and the R process or state reduction process which makes its appearance  whenever a
 measurement is invoked. The point is that without recourse to the R process
 the theory can make no predictions. But, when exactly should we in principle call upon the R process becomes a question that
 is not addressed within the theory. Other interpretations  have similar problems, for instance in the many worlds interpretation 
one has the universe splitting with every measurement. However the issue of how  in principle  do we determine what
 constitutes a measurement, is no resolved.
These issues have prompted  R. Penrose to propose that Quantum gravity might play a role in triggering a real
 dynamical collapse of the 
wave function of systems \cite{Penrose}.  His proposals would have a system collapsing whenever the gravitational  interaction 
energy between two alternative 
realizations that appear as superposed in a wave  function of a system reaches a certain threshold which is identified with
 $M_{Planck}$.
The
 ideas  can in principle lead to observable effects and in fact experiments to test them are currently being contemplated \cite{ExpPenrose}
(although it seems that the available technology can not yet be pushed to the level where actual tests might be expected to become a
 reality soon). We have considered in \cite{InflationUS} a situation for which  there exist already a wealth of empirical information and
 which we have argued
can not be fully understood without  involving some New Physics,  whose required  features would  seem to be quite
 close to Penrose's proposals:
  The quantum origin of the seeds of cosmic structure.

 In fact one of the major claimed  successes of Inflationary cosmology is its  reported ability 
to predict the correct spectrum for the primordial density fluctuations that seed the
growth of structure in our Universe. However 
when one thinks about it one immediately notes that there 
is something truly remarkable
 about it, namely that out of an initial situation which is taken to be  perfectly isotropic and homogeneous  
and based on a dynamics that preserves
 those symmetries one ends with a non-homogeneous and non isotropic situation.  Most of our colleagues who have been working in this 
field for a long time would reassure us,  that there is no problem at all by
invoking a variety of arguments. It is noteworthy  that these arguments  would tend to differ in general from one inflationary cosmologist
 to another \cite{Cosmologists}.  Other cosmologists do acknowledge that there seems to be something unclear at this point \cite{Cosmologists2}.
 In 
a recent paper \cite{InflationUS}
a critical analysis of such proposals has been carried out indicating that all the existing  justifications fail to be fully
 satisfactory.
 In particular, the cosmological situation  can be seen to be quite different from any other situation usually treated using
 quantum mechanics when one   notes the fact  that while in analyzing ordinary situations quantum  mechanics  offers us, at least one  
self consistent assignment 
at  all times of a state of the Hilbert space to our physical system (we are of course thinking of the Schroedinger picture). 
It is well known, that  in certain  instances there might be several 
mutually incompatible assignments of that sort,  as for instance  when contemplating the  two descriptions offered by two different inertial
 observers who
consider a given a specific EPR experiment.
 However,  as  we said, in all known cases, one has at least one description available. The reader might want to attempt to
 conceive  of  such
 assignment -- of a state at each time -- when presented  with any of the proposed  justifications offered to deal with the issue 
of the
 transition from a
 symmetric universe to a non-symmetric one. The reader will find that each instance he/she will be asked to accept one of the
 following: i)
our universe was not really 
in that symmetric state (corresponding to the vacuum of the quantum field), ii) our universe is still described by a symmetric state, 
iii) at least at some points in the past the description of the state of our universe could not be done within quantum mechanics, iv) 
quantum mechanics does not correspond to
the full description of  a system at all times,  or v) our own observations of the
 universe mark the transition from a symmetric to an asymmetric state. It should be clear that none of these  represent 
a satisfactory alternative,
in particular if we want to claim that we understand the  
evolution of our universe, its structure -- including ourselves -- , as the result of the fluctuations of quantum origin in 
the very early stages of 
our cosmology.
Needless is to say that
 none of these options will be explicitly called upon in the arguments one is presented with, however 
 one or more would be
 hidden, perhaps
 in a subtle way, underneath some of the aspects of 
the explanation. For a more thorough discussion we refer the reader to \cite{InflationUS}.

The interesting part of these situation is that one is forced to call upon to some novel physical
 process  to fill in  the
 missing or unacceptable part of the justification of the steps that are used to take us from that 
early and symmetric state, to the
 asymmetric state 
of our universe today, or  the state of the universe we photograph when we look at the surface of
 last scattering in the pictures of the CMB.
In \cite{InflationUS} we have considered in this  cosmological context a proposal calling for a 
self induced collapse of the wave function 
along the general 
lines conceived by Penrose, and
have shown that the requirement that one  should obtain results compatible with current observations
 is already sufficient
 to restrict 
in important 
ways some  specific aspects of these novel physics. Thus, when  we consider, that the origin of
 such new physics can be traced to some aspects of quantum gravity, one is already in a position of setting
 phenomenological
 constraints
at least on this aspect of the quantum 
theory of gravitation.  

In the following we give a short description of this analysis for the benefit of the reader.  The
staring point is as usual the action of a scalar field coupled to
gravity.
\be
\label{eq_action}
S=\int d^4x \sqrt{-g} \lbrack {1\over {16\pi G}} R[g] - 1/2\nabla_a\phi
\nabla_b\phi g^{ab} - V(\phi)\rbrack,
\ee
 where $\phi$ stands for the inflaton or scalar field responsible for inflation and $V$ for the 
inflaton's potential.
 One then splits both, metric and
scalar field into a spatially homogeneous (`background') part and an
inhomogeneous part (`fluctuation'), i.e. $g=g_0+\delta g$,
$\phi=\phi_0+\delta\phi$.

 The unperturbed solution correspond to the standard inflationary cosmology  which written using a conformal time,
 has a scale factor
\begin{equation}
a(\eta)=-\frac{1}{H_{\rm I} \eta},
\label{expansion}
\end{equation}
and with the scalar $\phi_0$ field in the slow roll regime.
The perturbed metric can be written
\begin{equation}
ds^2=a(\eta)^2\left[-(1+ 2 \Psi) d\eta^2 + (1- 2
\Psi)\delta_{ij} dx^idx^j\right],
\end{equation}
 where $\Psi$  stands for the relevant perturbation and is called
the Newtonian potential.

The perturbation of the scalar field leads to a perturbation of the energy momentum tensor, and
thus Einstein's equations at lowest order lead to
\begin{equation}
\nabla^2 \Psi  = 4\pi G \dot \phi_0 \delta\dot\phi .
\label{main2}
\end{equation}

Now, write the quantum theory of the field $\delta\phi$.
It is convenient  to consider instead the  field  $y=a \delta \phi$. 
We
consider the field in a box of side $L$, and  decompose the  real
field $y$  into plane waves
\begin{equation}
y(\eta,\vec{x})=\frac{1}{L^{3}} \Sigma_{ \vec k} \left(\ann_k y_k(\eta)
e^{i \vec{k}\cdot\vec{x}}+\cre_{k} \bar y_k(\eta)
e^{-i\vec{k}\cdot\vec{x}}\right),
\end{equation}
where the sum is over the wave vectors $\vec k$ satisfying $k_i L=
2\pi n_i$ for $i=1,2,3$ with $n_i$ integers. 

It is convenient to rewrite the field and momentum operators  as
\begin{equation}
\y(\eta,\vec{x})=
 \frac{1}{L^{3}}\sum_{\vec k}\ e^{i\vec{k}\cdot\vec{x}} \hat y_k
(\eta), \qquad \py(\eta,\vec{x}) =
\frac{1}{L^{3}}\sum_{\vec k}\ e^{i\vec{k}\cdot\vec{x}} \hat \pi_k
(\eta),
\end{equation}
where $\hat y_k (\eta) \equiv y_k(\eta) \ann_k +\bar y_k(\eta)
\cre_{-k}$ and  $\hat \pi_k (\eta) \equiv g_k(\eta) \ann_k + \bar g_{k}(\eta)
\cre_{-k}$
with
\begin{equation}
y^{(\pm)}_k(\eta)=\frac{1}{\sqrt{2k}}\left(1\pm\frac{i}{\eta
k}\right)\exp(\pm i k\eta),
\end{equation}
and
\begin{equation}
g^{\pm}_k(\eta)=\pm
i\sqrt{\frac{k}{2}}\exp(\pm i k\eta) . \label{Sol-g} 
\end{equation}

 As we will  be interested in considering a kind of self induced collapse which
 operates in close analogy with  a ``measurement", we proceed to work
 with  Hemitian operators, which in ordinary quantum mechanics are the ones susceptible of direct measurement.
Thus we decompose both $\hat y_k (\eta)$ and $\hat \pi_k
(\eta)$ into their real and imaginary parts $\hat y_k (\eta)=\hat y_k{}^R
(\eta) +i \hat y_k{}^I (\eta)$ and $\hat \pi_k (\eta) =\hat \pi_k{}^R
(\eta) +i \hat \pi_k{}^I (\eta)$ where
\begin{equation}
\hat{y_k}{}^{R,I} (\eta) =
\frac{1}{\sqrt{2}}\left(
 y_k(\eta) \ann_k{}^{R,I}
 +\bar y_k(\eta) \cre{}^{R,I}_k\right) ,\qquad  
\hat \pi_k{}^{R,I} (\eta) =\frac{1}{\sqrt{2}}\left( g_k(\eta)
\ann_k{}^{R,I}
 + \bar g_{k}(\eta) \cre {}^{R,I}_{k} \right).
\end{equation}
We note that the operators $\hat y_k^{R, I} (\eta)$ and $\hat
\pi_k^{R, I} (\eta)$ are therefore hermitian operators.  
Note that the operators corresponding to $k$ and $-k$ are identical in the real
case (and identical up to a sign in the imaginary case).

Next we specify our model of collapse, and follow the field evolution through collapse
to the end of inflation.
  We will assume that the collapse is
somehow analogous to an imprecise measurement of the
operators $\hat y_k^{R, I}
(\eta)$ and $\hat \pi_k^{R, I} (\eta)$ which, as we pointed out are
hermitian operators and thus reasonable observables.  These field
operators contain complete information about
the field (we ignore here for simplicity the relations between the modes $k$ and $-k$).

 Let $|\Xi\rangle$ be any state in the Fock space of
$\hat{y}$. Let us introduce the following quantity:
$
 d_k^{R,I} = \l \ann_k^{R,I} \r_\Xi.
$
Thus  the expectation values of the modes are expressible
as
\begin{equation}
\l {\y_k{}^{R,I}} \r_\Xi = \sqrt{2} \Re (y_k d_k^{R,I}),  \qquad
\l {\py_k{}^{R,I}} \r_\Xi = \sqrt{2} \Re (g_k d_k^{R,I}).
\end{equation}

For the vacuum state $|0\rangle$ we  have of course:
$
\l{\y_k{}^{R,I}}\r_0 = 0, \l\py_k{}^{R,I}\r_0 =0,
$
while their corresponding uncertainties are
\begin{equation}\label{momentito}
\fluc{\y_k {}^{R,I}}_0 =(1/2) |{y_k}|^2(\hbar L^3), \qquad
\fluc{\pf_k {}^{R,I}}_0 =(1/2)|{g_k}|^2(\hbar L^3).
\end{equation}

{\bf The collapse}\newline

Now we will specify the rules according to which collapse happens.
Again, at this point our criteria will be simplicity and naturalness.
Other possibilities do exist, and may lead to different
  predictions. 

What we have to describe is the state $|\Theta\rangle$ after the
collapse. We need to specify 
$d^{R,I}_{k} = \langle\Theta|\ann_k^{R,I}|\Theta\rangle $
In the vacuum state, $\y_k$ and
$\py_k$ individually are distributed according to Gaussian
distributions centered at 0 with spread $\fluc{\y_k}_0$ and
$\fluc{\py_k}_0$ respectively.  However, since they are mutually
non-commuting, their distributions are certainly not independent.  In
our collapse model, we do not want to distinguish one over the other,
so we will ignore the non-commutativity  and make the following
assumption about the (distribution of) state(s) $|\Theta\rangle$ after
collapse:
\begin{eqnarray}
\l {\y_k^{R,I}(\eta^c_k)} \r_\Theta&=&x^{R,I}_{k,1}
\sqrt{\fluc{\y^{R,I}_k}_0}=x^{R,I}_{k,1}|y_k(\eta^c_k)|\sqrt{\hbar L^3/2},\\
\l {\py_k{}^{R,I}(\eta^c_k)}\r_\Theta&=&x^{R,I}_{k,2}\sqrt{\fluc{\pyRI_k}
_0}=x^{R,I}_{k,2}|g_k(\eta^c_k)|\sqrt{\hbar L^3/2},
\end{eqnarray}
where $x_{k,1},x_{k,2}$ are distributed according to a Gaussian
distribution centered at zero with spread one.
From these equations we  solve for $d^{R,I}_k$.
 Here we must recognize that our universe, corresponds 
to a single realization of the random variables, and thus  each of the quantities 
$ x^{R,I}{}_{k,1,2}$ has a  single specific value. 
Latter we will se how to make relatively specific predictions despite of these features.

Next we focus on the expectation value of the quantum
operator which appears in our basic formula
\begin{equation}\nabla^2 \Psi = s \Gamma \label{main3}
\end{equation}
(where we introduced the abbreviation $s=4\pi G \dot \phi_0$) and the
quantity $\Gamma$ as the aspect of the field that acts as a source of
the Newtonian Potential.  In the slow roll approximation we have
$\Gamma=\delta\dot\phi= a^{-1} \pi^{y}$. We want to say that, upon
quantization, the above equation turns into
\begin{equation}\nabla^2 \Psi = s \langle\hat\Gamma\rangle. \label{main4}
\end{equation}
Before the collapse occurs, the expectation value on the right hand
side is zero. Let us now determine what happens after the collapse: To
this end, take the Fourier transform of (\ref{main4}) and rewrite it
as
\begin{equation}\label{modito}
\Psi_k(\eta)=\frac{s}{k^2}\langle\hat\Gamma_k\rangle_\Theta.
\label{Psi}
\end{equation}

Let us focus now on the slow roll approximation and compute the right
hand side, we note that $\delta\dot\phi=a^{-1}\py$ and hence
 we find
\begin{eqnarray}
\nonumber
\langle\Gamma_k\rangle_\Theta&=&\sqrt{\hbar L^3 k}\frac{1}{2a}F(k), \label{F}
\end{eqnarray}
where
\begin{equation}
F(k) = (1/2) [A_k (x^{R}_{k,1} +ix^{I}_{k,1}) + B_k (x^{R}_{k,2}
+ix^{I}_{k,2})],
\end{equation}
with
\begin{equation}  A_k =  \frac {\sqrt{ 1+z_k^2}} {z_k} \sin(\Delta_k) ; \qquad  B_k
=\cos (\Delta_k) + (1/z_k) \sin(\Delta_k)
\end{equation}
and where  $\Delta_k= k \eta -z_k$ with $ z_k =\eta_k^c
k$.

  Next we turn to the  experimental  results. We will for the most part,  disregard the changes to
dynamics that happen after re-heating  and due to the transition to
standard (radiation dominated) evolution.
The  quantity that is measured is ${\Delta T \over T}
(\theta,\varphi)$ which is a function of the coordinates on the
celestial two-sphere which is expressed as $\sum_{lm} \alpha_{lm}
Y_{l,m}(\theta,\varphi)$.  The angular variations of the 
temperature are then identified with the corresponding variations in the
``Newtonian Potential" $ \Psi$, by the understanding that they are the
result of gravitational red-shift in the CMB photon frequency $\nu$ so
${{\delta T}\over T}={{\delta \nu}\over {\nu}} = {{\delta (
    \sqrt{g_{00}})}\over {\sqrt{g_{00}}}} \approx\delta \Psi$.  

 The quantity that is presented
as the result of observations is $OB_l=l(l+1)C_l$ where $C_l =
(2l+1)^{-1}\sum_m |\alpha^{obs}_{lm}|^2 $. The observations indicate
that (ignoring the acoustic oscillations, which is anyway an aspect
that is not being considered in this work) the quantity $OB_l$ is
essentially independent of $l$ and this is interpreted as a reflection
of the ``scale invariance" of the primordial spectrum of fluctuations.

Then, as we noted the  measured quantity is the
``Newtonian potential" on the surface of last scattering: $
\Psi(\eta_D,\vec{x}_D)$,  from where one 
extracts
\begin{equation}
\a_{lm}=\int \Psi(\eta_D,\vec{x}_D) Y_{lm}^* d^2\Omega.
\end{equation}
To evaluate the  expected value for the quantity of interest we use (\ref{Psi}) and (\ref{F}) to
write
\begin{equation}
 \Psi(\eta,\vec{x})=\sum_{\vec k}\frac{s  U(k)} {k^2}\sqrt{\frac{\hbar
k}{L^3}}\frac{1}{2a}
 F(\vec{k})e^{i\vec{k}\cdot\vec{x}},
\label{Psi2}
\end{equation}
where we have added the factor $U(k)$ to represent the aspects of
the evolution of the quantity of interest associated with the
physics of period from re-heating to de coupling, which includes among
  others the acoustic oscillations of the plasma.  

 After some algebra we obtain
\begin{eqnarray}
\alpha_{lm}&=&s\sqrt{\frac{\hbar}{L^3}}\frac{1}{2a} \sum_{\vec
k}\frac{U(k)\sqrt{k}}{k^2} F(\vec k)  4 \pi i^l  j_l((|\vec k|
R_D) Y_{lm}(\hat k),\label{alm1}
\end{eqnarray}
where $\hat k$ indicates the direction of the vector $\vec  k$. It is in this
expression that the justification for the use of statistics becomes
clear.  The quantity we are in fact considering is the result of 
the combined contributions of an
ensemble of harmonic oscillators each one contributing with a complex
number to the sum, leading to what is in effect a 2 dimensional random
walk whose total displacement corresponds to the observational
quantity. To proceed further we must evaluate the most likely value
for such total displacement. This we do with the help of the imaginary
ensemble of universes, and the identification of the most likely value
with the ensemble mean vale. Now we
compute the expected magnitude of this quantity. After taking the continuum limit we find, 
\begin{equation}
|\alpha_{lm}|^2_{M. L.} 
=\frac{s^2  \hbar}{2 \pi a^2} \int \frac {U(k)^2
C(k)}{k^4} j^2_l((|\vec k| R_D)  k^3dk, \label{alm4}
\end{equation}
where  
\begin{equation}
C(k)=1+ (2/ z_k^2) \sin (\Delta_k)^2 + (1/z_k)\sin (2\Delta_k).
\label{ExpCk}
\end{equation}

  The last expression can be made more useful
by changing the variables of integration to $x =kR_D$ leading to
\begin{equation}
|\alpha_{lm}|^2_{M. L.}=\frac{s^2   \hbar}{2 \pi a^2} \int
\frac{U(x/R_D)^2 C(x/R_D)}{x^4}    j^2_l(x) x^3 dx,
\label{alm5}
\end{equation}
which in the exponential expansion regime where $\mu$ vanishes and in
the limit $z_k\to -\infty$ where $C=1$, and taking for simplicity
  $U (k) =U_0$ to be independent of $k$, (neglecting for instance the
  physics that gives rise to the acoustic peaks), we find:
 \begin{equation}
 |\alpha_{lm}|^2_{M. L.}=\frac{s^2  U_0^2  \hbar} {2  a^2}
\frac{1}{l(l+1)} .
\end{equation}

 Now,  since this does not depend on $m$ it
is clear that the expectation of $C_l = (2l+1)^{-1}\sum_m
|\alpha_{lm}|^2 $ is just $|\alpha_{lm}|^2$ and thus the observational
quantity $OB_l=l(l+1)C_l =\frac{s^2 U_0^2 \hbar}{2 a^2} $ independent
of $l$ and in agreement with the scale invariant spectrum obtained in
ordinary treatments and in the observational studies.  

Thus, the predicted value for the  $OB_l$ is \cite{InflationUS},
\begin{equation}
OB_l= (\pi/6) G\hbar \frac{(V')^2}{V} U_0^2 =
(\pi/3)\epsilon (V/M_{Pl}^4) U_0^2,
\end{equation}
where  we have used the standard definition of the
slow roll parameter $\epsilon= (1/2) M_{Pl}^2 (V'/V)^2$ which is  normally expected to be rather small. 
 We note that if one could avoid $U$ from becoming
too large during re-heating, the quantity of interest
  would be proportional to $\epsilon$ a possibility
 that was not uncovered  in the standard treatments, so one 
  could get rid of the ``fine tuning problem" for the inflationary
  potential, i.e. even if $ V\sim M_{Pl}^4$, the temperature
  fluctuations in the CMB would be expected to be small.

Now let us focus on the effect of the finite value of times of
collapse $\eta^c_k$, that is, we consider the general functional form of
$C(k)$. The first thing we note is that in order to get a reasonable
spectrum there seems to be only one simple option: That $z_k $ be
essentially independent of $k$ that is the time of collapse of the
different modes should depend on the mode's frequency according to
$\eta_k^c=z/k$. This is a remarkable
conclusion which would  provide relevant information about whatever the
mechanism of collapse is. 

 Lets turn next to one simple proposal about the collapse mechanism which following Penrose's ideas is assumed to
be tied to Quantum Gravity, and examine it with the above results in mind.
%---------------------------------------------------------
\subsection{ A version of  `Penrose's  mechanism' for collapse in the cosmological  setting}
\label{sec_penrose}
%--------------------------------------------------------
Penrose has for a long time advocated  that  the collapse of quantum
mechanical wave functions might be a dynamical process independent of observation, and that the
underlying mechanism might be related to gravitational
interaction. More precisely, according to this suggestion, collapse
into one of two quantum
mechanical alternatives would take place when the gravitational
interaction energy between the alternatives exceeds a certain
threshold. In fact, much of the initial motivation for the present
work came from Penrose's ideas and his questions regarding the quantum
history of the universe.

A very naive realization of Penrose's ideas in the present setting
could be obtained as follows: Each mode would collapse by the
action of the gravitational interaction between it's own possible
realizations. In our case one could estimate the interaction energy
$E_I(k,\eta)$ by considering two representatives of the possible
collapsed states on opposite sides of the Gaussian associated with
the vacuum. Let us interpret $\Psi$ literally as the Newtonian
potential and consequently the right hand side of equation
(\ref{main2}) as the associated matter density $\rho$.  Therefore, $\rho
=\dot\phi_0 \Gamma $, with $\Gamma =\pi^y/a$. Then we would have:
\be 
E_I(\eta)=\int \Psi^{(1)}(x,\eta) \rho^{(2)}(x,\eta)dV = a^3
\int \Psi^{(1)}(x,\eta) \rho^{(2)}(x,\eta) d^3x, 
\ee 
which when
applied to a single mode becomes: 
\be E(\eta)= (a^3/L^6)  \Psi_{
k}^{(1)}( \eta) \rho^{(2)}_{k} (\eta) \int d^3x = (a^3/L^3)
\Psi^{(1)}_{ k}( \eta) \rho^{(2)}_{k} (\eta), 
\ee 
where $(1),(2)$
refer to the two different realizations chosen. Recalling
 that $\Psi_{ k} = ( s/k^2) \Gamma_k$, with $s= 4\pi G\dot\phi_0$, and using equation (\ref{momentito}), we get
 $|<\Gamma_k > |^2 = \hbar k L^3 (1/2a)^2$.
Then 
\be
 E_I(k,\eta) = ( \pi/4) (a/k) \hbar G (\dot\phi_0)^2.
 \ee
In accordance to Penrose's ideas the collapse would take place
when this energy reaches the `one-graviton' level, namely when
$E_I(k,\eta)=M_p$, where $M_p$ is the
Planck mass, thus 
one gets $ z_k=\frac{\pi \hbar G \dot \phi_0^2}{H_I M_p}$. So
$z_k$ is independent of $k$ which  leads to a roughly  scale invariant spectrum of fluctuations in
accordance with observations.
Thus a naive realization of Penrose's ideas seems to be a good candidate to supply the element that we argued  
is missing in the standard accounts of the emergence of the seeds of cosmic structure from quantum fluctuations
 during the inflationary regime in the early universe.

\section{Conclusions}

The dramatic change in outlook that has taken place in the last few years regarding the possibility
 -- despite early pessimistic assessments -- that some aspects of quantum gravity might be
 after all experimentally accessible is a very healthy 
development for the quantum gravity community. Bringing back to the realm of empirical falsificability 
of ideas, a discipline 
that seemed to wander ever deeper into the abyss of unchecked  lucubrations, can not but reassure us, 
that the discipline
 still lies within the boundaries of  scientific research. The  early proposals that there might be a 
breakdown in Lorentz 
Violation associated with a  discrete structure that quantum gravity is supposed
 to endow space-time with, lead in fact to a vigorous program, which at this time,
and due  not only to the  direct bounds obtained but more importantly to the severe restrictions that
 QFT puts on these ideas,
 have to be regarded  with a strong dose of skepticism, as the lesson from that stage seem to be in
 the direction
  of requiring Quantum Gravity to be free of such effects \footnote{ There seems to be at this time a 
single situation
where some indications that a breakdown of Lorentz invariance could be at play: The  absence of a GZK 
cut-off in the cosmic ray
 spectrum \cite{GZK}.
 The evidence is  still rather controversial \cite{GZKN}
 and, on the other hand it is not clear whether some  simpler explanations, perhaps including new physics,
 but unconnected with the issues at hand, do exist.}.  The lasting legacy  of this episode, on the other hand,
 is, I believe, 
the lesson that we 
should not give up so easily in our quest for phenomenological manifestations of quantum gravity. 
 In this regard 
 the successors of the program should be divided in three groups, first those ideas that suffer from 
a lack of clear interpretational status including some  for which the existence of a sensible
 interpretational scheme is highly dubious, and which are
briefly discussed in section 2, then there are some
 ideas that seem to be well defined and have a rather clear interpretational status, and which could,
 in principle be subjected to experimental 
investigations, such as the specific search for a gravitationally induced collapse of the wave
 function proposed by Penrose \cite{Penrose}, 
 the proposals of  Pullin and Gambini about a  gravitationally induced  fundamental 
decoherence  \cite{GIDecoh}, the ideas about
 a possible non-standard manifestations of curvature in extended quantum systems,
 first proposed in \cite{NewQGP} and reviewed in  section 3.  Finally,  as was described in section 4, 
and first reported in \cite{InflationUS}
the recognition that there are very 
intriguing aspects of our understanding of the origin of the seeds of cosmic structure, which seem 
 to ``account" for the observations, in the sense that the predictions and observations are in agreement,
 but that on the other hand suffer from unjustified identifications, problematic  interpretations, 
and do not pass a
 careful and profound examination. In other  words  the recognition that something else seems to be needed 
for the whole picture to work, could be pointing us towards an actual manifestation quantum gravity. 
We have shown that not only 
the issues are susceptible of scientific investigation based on observations, but that a simple
 account of what is needed seem to be
 provided by the extrapolation of Penrose's ideas to the cosmological setting.

We end by stressing that it might well be that we are at the dawn of a new era regarding Quantum Gravity;
 but we would do well by  keeping an open mind, as it is quite likely that such new era, as any  region which 
is truly virgin to exploration, will look rather different that what was expected on arrival.

\section*{Acknowledgments}

\noindent   It is a pleasure to acknowledge very  helpful conversations with Chryssomalis 
 Chryssomalakos. This work was supported
 in part  by DGAPA-UNAM
IN108103 and CONACyT 43914-F grants.

\end{document}